\documentclass[traditabstract]{aa} % for tradit abstract

\usepackage{latexsym}
\usepackage{amssymb}

\usepackage{graphicx}
\usepackage{txfonts}
\title{65 kpc of ionized gas trailing behind NGC 4848 during its first crossing of the Coma cluster
\thanks{Observations taken at the  observatory of San Pedro Martir (Baja California, Mexico), belonging to the
Mexican Observatorio Astron\'omico Nacional.}
}

\author{Matteo Fossati \inst{1}
\and Giuseppe Gavazzi \inst{1}
\and Alessandro Boselli \inst{2}
\and Michele Fumagalli \inst{3}
}

\authorrunning{M. Fossati et al.}
\titlerunning{$\rm H\alpha$ tail behind NGC 4848}

\institute{Universit\`a degli Studi di Milano-Bicocca, Piazza della Scienza 3, 20126 Milano, Italy\\
\email {matteo.fossati@mib.infn.it; giuseppe.gavazzi@mib.infn.it}
\and
Laboratoire d'Astrophysique de Marseille, UMR 6110 CNRS, 38 rue F. Joliot-Curie, F-13388, Marseille, France\\
\email {alessandro.boselli@oamp.fr}
\and
Department of Astronomy and Astrophysics, University of California, 1156 High Street, Santa Cruz, CA 95064, USA\\
\email {mfumagalli@ucolick.org}
}

\begin{document}
\date{Received July 2, 2012; accepted July 18, 2012}

% 5 {} token are mandatory

\abstract
{In a five hour $\rm H\alpha$ exposure 
of the northwest region of the Coma cluster with the 2.1m telescope at San Pedro
Martir (Mx),
we discovered a 65 kpc cometary emission of ionized gas trailing behind the SBab galaxy NGC 4848.
The tail points in the opposite direction of the cluster center, in the same direction
where stripped HI had been detected in previous observations.
The galaxy  shows bright HII regions in an inner ring-like pattern, 
where the star formation takes place at the prodigious rate of $\rm \sim 8.9~ M_{\odot}~ yr^{-1}$.
From the morphologies of the galaxy and the trailing material, we infer that the galaxy
is suffering from ram pressure due to its high velocity motion through the
intergalactic medium.
We estimate that $\sim 4 \times 10^9$ $\rm M_{\odot}$ of gas is swept out from the galaxy forming the tail.
Given the ambient conditions in the Coma cluster ($\rho_0=6.3\times 10^{-27} ~\rm g~ cm^{-3}$;
$\sigma_{vel}=940 ~\rm km~s^{-1}$), simulations predict  that the ram pressure mechanism is able to remove such an amount of gas 
 in less than 200 Myr. This, combined with the geometry of the interaction, is
 indicative of radial infall into the cluster,
 leading to the conclusion that  NGC 4848 has been caught during its first passage through the dense cluster environment.
}

  %\abstract
      

\maketitle														     	

%
%________________________________________________________________

\section{Introduction}

Yagi et al. (2010) reported a deep (4.5 hour integration) H$\alpha$ survey covering the central $0.5\times 0.5 \rm ~deg^2$ of the Coma cluster (A1656) with the Suprime-Cam mounted on the Subaru telescope. 
Unexpectedly for an evolved cluster such as Coma, these authors found that almost every star-forming member
has its own spectacular complex of diffuse, ionized, gaseous trails extending dozens of kpc behind the optical extent of the galaxies, 
and sometimes harboring star-forming compact knots. They revealed 14 such systems,
including those previously reported by Yagi et al. (2007) and Yoshida et al. (2008) in the Coma cluster. 
Similar examples can also be found in A1367 (Gavazzi et al. 2001),
in Virgo (Yoshida et al. 2002, 2004; Kenney et al. 2008), and in A3627 (Sun et al. 2007).
These features suggest that the galaxies were recently captured by the cluster gravitational potential and are
now infalling toward the cluster center (Yagi et al. 2010).
Unfortunately, the field of view of the Suprime-Cam missed by less than 5 arcmin the position of NGC 4848
which is another obvious 
candidate for possible extended $\rm H\alpha$ emission. Bothum \& Dressler (1986) had indeed listed 
NGC 4848 as one of the dozen unusually active galaxies found in the Coma cluster.

\object{NGC 4848} (CGCG 160-055; Zwicky et al. 1961-68) is a bright ($M_B$=-20.5) SBab:edge-on (RC3, de Vaucouleurs et al. 1991)
galaxy that lies at the northwest (N-W) periphery of the X-ray emitting region in the Coma cluster. 
It has a vigorous star-formation rate of $\rm \sim 9 ~M_{\odot} ~yr^{-1}$  as derived from
the $\rm H\alpha$, ultraviolet (UV), far-infrared (FIR), and radio-continuum emission (see \S \ref{galaxy}).
Observations in the 21 cm line of HI (Gavazzi 1989; Bravo Alfaro et al. 2001) revealed
a moderately deficient HI content ($Def_{\rm HI}=0.46$), displaced in the N-W direction, 
as opposed to its H$_2$ content, which appears normal and centrally concentrated (Vollmer et al. 2001). 
The asymmetry in the HI distribution suggests that the galaxy is experiencing ram pressure (Gunn \& Gott 1972) 
owing to its high velocity motion through the intergalactic medium (IGM). 
This discrepancy is expected since unless ram pressure stripping is severe, 
only the atomic phase of the gas distributed at the galaxy periphery is removed, while the $\rm H_2$,
bound deep within the galaxy potential well, is mostly unaffected by ram pressure (Combes et al. 1988;
Kenney \& Young 1989; Boselli et al. 2002; Fumagalli \& Gavazzi 2008).

Numerical hydrodynamical simulations of galaxies subject to ram pressure stripping in rich clusters
(e.g. Kapferer et al. 2009; Tonnesen \& Bryan 2009, 2010, 2012; Ruszkowski et al. 2012) 
reveal that in much less than 1 Gyr these galaxies lose
all of their gas when the density of the
IGM and the transit velocity are as high as in the Coma cluster. 
Consistent results are found both with and without magnetic fields.
Extended gaseous tails form and the gas is shocked and heated by turbulence (Yoshida et al. 2004, 2012; 
Kenney et al. 2008), producing compact knots where radiative cooling takes place favoring the 
star formation.

\begin{table*}[!t]
\begin{center}
\caption{Log book of the imaging observations.
%The imaging instrumental set-up
}
\label{Tab1}
\begin{tabular}{cccccccc}
\hline
\hline
\noalign{\smallskip}
Telescope & Date & CCD & Pix  & Filter  & Tint & Nexp &seeing\\
           &        &       &  (arcsec) & (\AA)     & (sec)  &      & (arcsec)\\
\noalign{\smallskip}
\hline
\noalign{\smallskip}
INT  & 20 Mar 1999 & $4 \times 2048 \times4100$ EEV  & 0.33 & B           & 300   & 2  & 1.0\\
INT  & 28 Apr 2000 & $4 \times 2048 \times4100$ EEV  & 0.33 & 6725 (80)   & 1200  & 3  & 1.3\\
INT  & 28 Apr 2000 & $4 \times 2048 \times4100$ EEV  & 0.33 & (Gunn) $r'$ & 300   & 3  & 1.3\\
TNG  & 09 Feb 2001 & $1024 \times 1024$ NICS         & 0.25 & H           & 60    & 9  & 0.8\\
SPM  & Apr 2012    & $1024 \times 1024$ EEV          & 0.35 & 6723 (80)   & 600   & 30 & 1.4\\
SPM  & Apr 2012    & $1024 \times 1024$ EEV          & 0.35 & (Gunn) $r$  & 60    & 26 & 1.4\\
\noalign{\smallskip}
\hline
\hline
\end{tabular}
\end{center}
\end{table*}

In 2000, we serendipitously discovered
a low surface-brightness $\rm H\alpha$ emission trailing behind NGC 4848 
(not reported by Iglesias-P{\'a}ramo et al. 2002) 
in a one-hour exposure of the central $1\times 1 \rm ~deg^2$ of the Coma cluster with the Wide Field Camera (WFC) 
at the Isaac Newton Telescope (INT, La Palma).
However, this extended emission was only marginally detected and follow-up observations were required.
Similar extended features were detected in deep GALEX images by Smith et al. (2010) showing several knots
of recent star-formation along the tail.
In 2012, we acquired additional five-hour observations with the San Pedro Martir 
(SPM) telescope using narrow-band $\rm H\alpha$ filters.
The resulting stacked six-hour exposure, which we present in this work,
is sufficiently deep to allow a robust determination of the flux in the tail.
Throughout this paper, we assume $\rm H_0=73~km~s^{-1}~Mpc^{-1}$, thus NGC 4848 is at the distance 
of 95.5 Mpc, that of the Coma cluster.

\section{Observations}

We observed  NGC 4848 in the B band using the WFC at the prime focus of the 2.5m INT in 1999 and 
in the H band using the Near Infrared Camera Spectrometer (NICS) mounted at the 3.6m Telescopio Nazionale Galileo (TNG) in 2001. 
Both observations (available through GoldMine, Gavazzi et al. 2003) 
were obtained under $\leq 1.0$ arcsec seeing (see Table \ref{Tab1}).
We also observed the field centered on NGC 4848 in $\rm H\alpha$ using two
telescopes: the INT, and the 2.1 m telescope at SPM.
The observations were performed through
narrow-band filters centered at $\sim$ 6725~\AA,
covering the redshifted $\rm H\alpha$ and [NII] lines (ON-band image).
The underlying continuum was measured through the broad-band $r$-Gunn filter 
(OFF-band image; see Table \ref{Tab1} for details).
The images were obtained in photometric conditions
with seeing ranging from 1.3 to 1.4 arcsec.
To minimize the impact of cosmic rays, we acquired a set of shorter exposures that were
subsequently  combined using a median filter.
Images of the spectrophotometric star Feige 34 were also collected for the photometric calibration.

The individual images were bias-subtracted and flat-fielded
using combinations of exposures of several empty fields at twilight.
A second-order flat-field correction was achieved using {\it superflat} images obtained by median-combining 
several science exposures that were 
selected because these are the ones containing only point-like objects 
among all exposures taken during the 2012 run.
The images were aligned using field stars as reference (using the {\sc IRAF} task {\it imalign}) 
and fitted individually with third-order polynomials to
correct for residual flat-field inhomogeneities (using the {\sc IRAF} task \emph{imsurfit}).
After background subtraction and rebinning the two sets of images to a common pixel scale of
0.35 $\rm arcsec~ pixel^{-1}$ (using the {\sc IRAF} task {\it imlintran}), we obtained the final combined image 
(using the {\sc IRAF} task {\it imcombine}).
The resulting stacked frames correspond in total to a 6 hour and 41 min integration for the ON-band and 
OFF-band, respectively. 

\begin{figure*}
\centering
\includegraphics[scale=0.24]{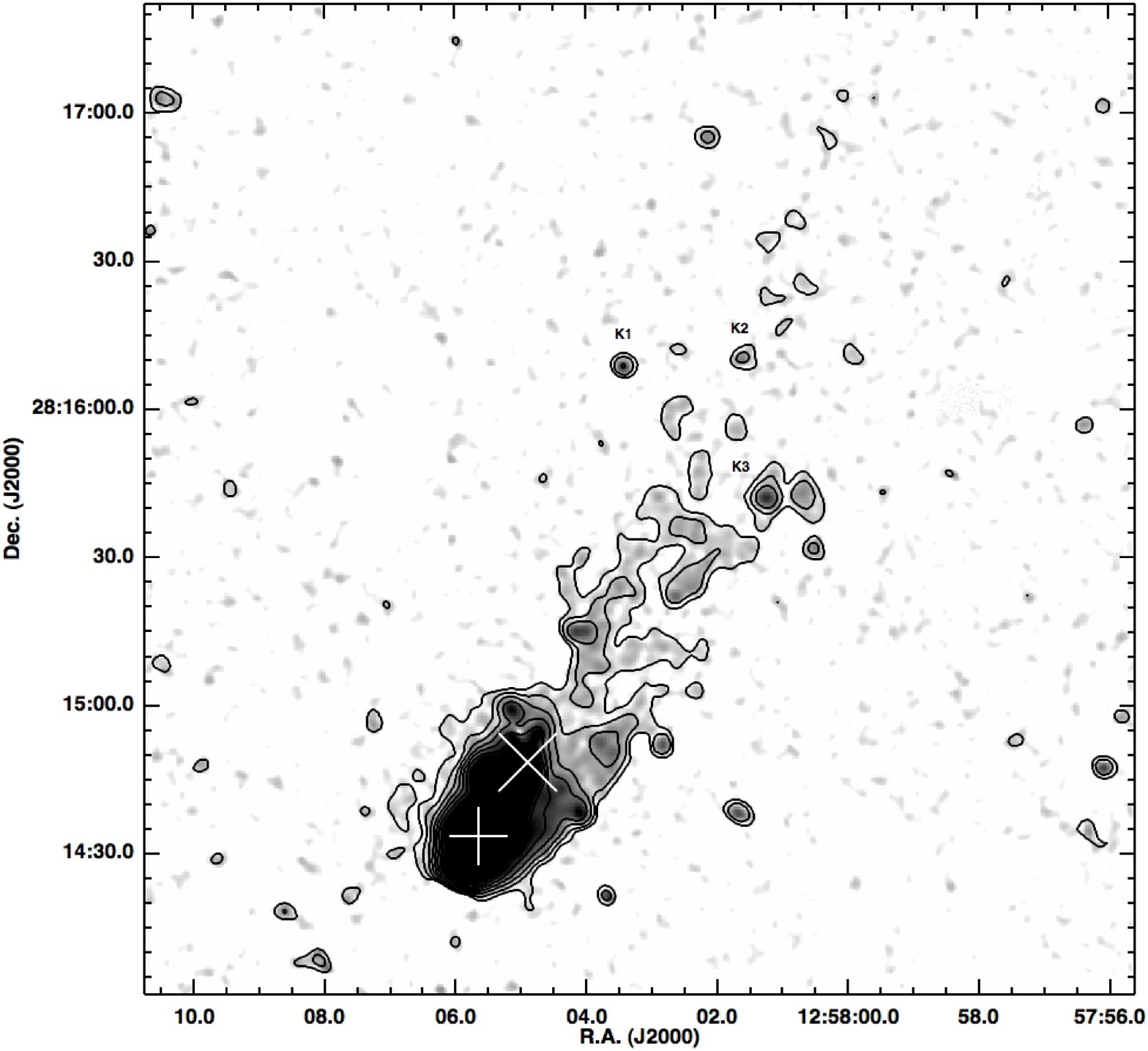}
\includegraphics[scale=0.24]{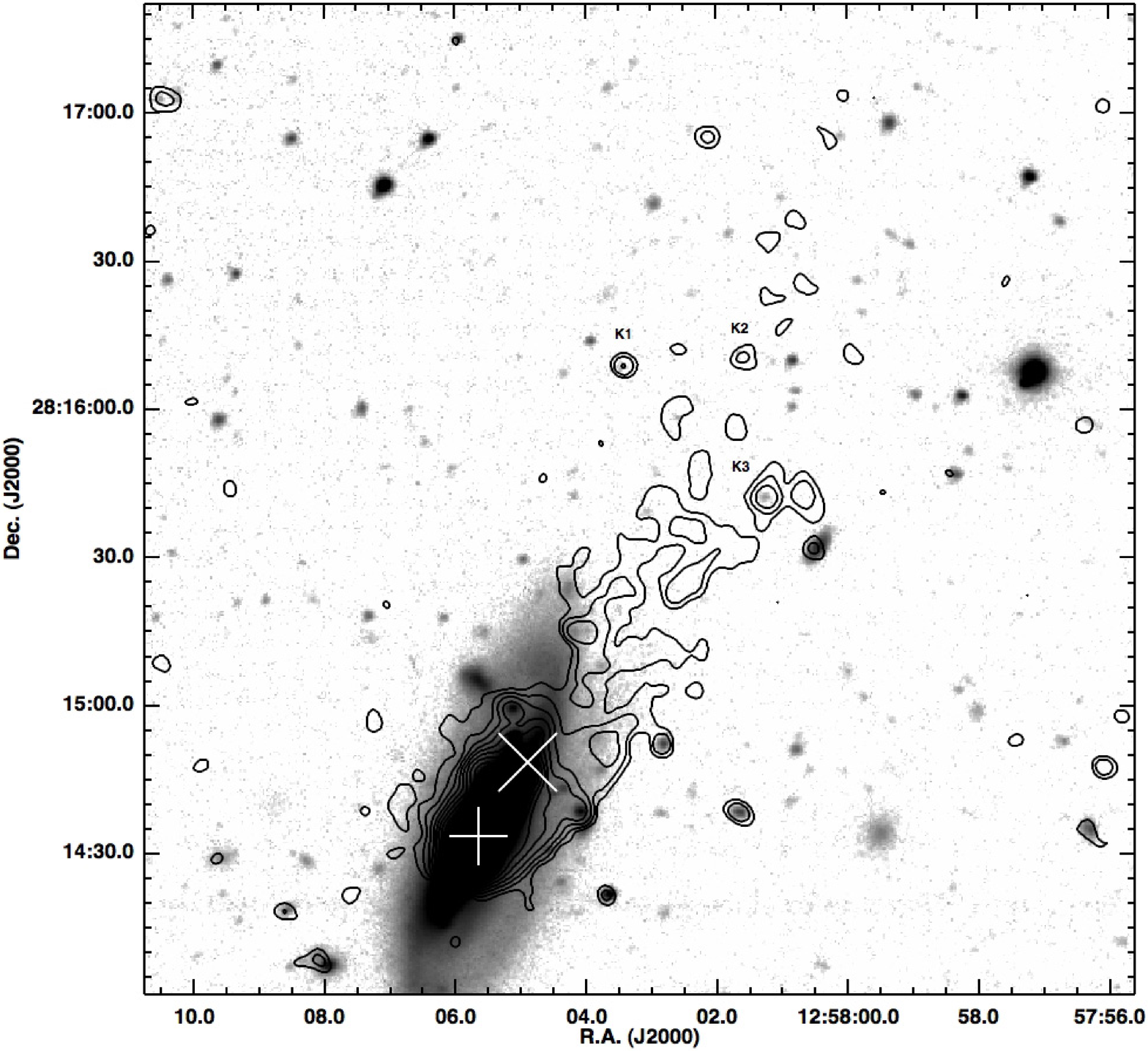}
\caption{Left panel: Grayscale representation of the Gaussian-smoothed (three-pixel-kernel) NET image of NGC 4848, with contours of the same image
(contours are from  1 to 320 $\times 10^{-17}~\rm erg~cm^{-2}~sec^{-1} arcsec^{-2}$  in 10 logarithmic intervals). The white cross marks the center 
of the molecular hydrogen given by Vollmer et al. (2001), while the white X marks the center of the HI given by Bravo Alfaro et al. (2001). Three knots of NET emission with associated
stellar continuum are marked K1-3.
Right panel: Grayscale representation of the OFF image of NGC 4848, with contours of the Gaussian-smoothed
(three-pixel-kernel) NET image (same as left panel). Note the angle between the major axis of the galaxy and the direction marked by the tail,
which points toward the center of the Coma cluster.}
\label{NET}
\end{figure*}

Finally, we derive the continuum-subtracted $\rm H\alpha$+[NII] image in the following way.
First, the intensity in the combined OFF-band image was normalized to that of the combined 
ON-band one, using the flux ratio of several field stars. 
Then, since Spector et al. (2012) pointed out that the colors of foreground stars differing from those 
in the target galaxies can be an important source of error when estimating the normalization coefficient, 
we applied a correction as follows.
We evaluated that the average color of NGC 4848 is $g-r_{gal} = 0.7$ mag (see Table \ref{Table1}).  
This is only slightly different from the mean color of Milky Way stars at similar Galactic latitude 
$g-r_{stars} = 0.8$ mag (Spector et al. 2012). 
From the linear fit shown in figure 2 of Spector et al. (2012), we infer that 
the normalization coefficient for NGC 4848 is 0.98 times the one estimated from the foreground stars.
In our case, this correction is only marginal. 
The combined NET-image was then obtained by subtracting the
normalized OFF-band frame from the ON-band one. The resulting OFF-band and 
NET frames are shown in Fig. \ref{NET}. 
The limiting $\rm H\alpha$ surface brightness was estimated using photometry of randomly sampled 
apertures of 8 arcsec in diameter. 
The resulting 1$\sigma$ limiting sensitivity is 
$1.0 \times 10^{-17}~\rm erg~cm^{-2}~sec^{-1} arcsec^{-2}$
(at full resolution), which is a factor of 
four shallower than the superb Subaru data from Yagi et al. (2010).

\begin{table}[ht!]
\caption{Physical parameters of NGC 4848 
that complement Table 1 of Vollmer et al. (2001).}
\centering
\begin{tabular}{ll}
\hline
\hline
$\alpha$ (J2000)          	  &   $12^h58^m05^s.6$      \\  
$\delta$ (J2000)          	  &   $28^\circ14'34"$      \\ 
Distance \tablefootmark{a}     	  &   95.5 Mpc \\ 
$u$\tablefootmark{b}	  	  &   15.82 mag\\
$g$\tablefootmark{b}	  	  &   14.54 mag\\
$r$\tablefootmark{b}	  	  &   13.81 mag\\
$i$\tablefootmark{b}	  	  &   13.42 mag\\
$z$\tablefootmark{b}	  	  &   13.17 mag\\
$ M_{\rm Dyn}$\tablefootmark{c}    &   $3.8\times 10^{11}~\rm M_{\odot}$	 \\
$ M_{\rm Star}$\tablefootmark{d}   &   $4.8\times 10^{10}~\rm M_{\odot}$	 \\  
$ M_{\rm HI}$\tablefootmark{e}     &   $2.2\times 10^{9}~\rm M_{\odot}$ 	\\
$ M_{\rm H_2}$\tablefootmark{f}    &   $4.1\times 10^{9}~\rm M_{\odot}$ 	\\
$ M_{\rm trail}$\tablefootmark{g}  &   $3.6\times 10^{9}~\rm M_{\odot}$ 	\\  
$ M_{\rm Bar}$\tablefootmark{h}    &   $5.8\times 10^{10}~\rm M_{\odot}$	 \\ 
$ M_{\rm Dyn}/M_{\rm Bar}$		  &   6.5      \\
$SFR$\tablefootmark{g}        &   $\rm  8.9 ~M_{\odot} yr^{-1}$  \\
\hline
\hline
\end{tabular}
\tablefoot{
\tablefoottext{a}{Assuming a distance modulus $m-M = 34.9~\rm mag$ (Gavazzi et al. 2010).}
\tablefoottext{b}{Petrosian AB magnitudes from the SDSS DR7 database (Abazajian et al. 2009). 
Corrected for extinction in the Galaxy (Schlegel et al. 1998).}
\tablefoottext{c}{Assuming centrifugal equilibrium, $V_{\rm rot}=270~\rm km~s^{-1}$ (Amram et al. 1992) 
 and the B-band diameter $a_{25}=1.6$ arcmin. (RC3, de Vaucouleurs et al. 1991)}
\tablefoottext{d}{Derived from the $i$ magnitudes and $g-i$ color using a
modified Bell et al. (2003) recipe consistent with the mass determination of MPA-JHU.}
\tablefoottext{e}{Gavazzi (1989)}
\tablefoottext{f}{Vollmer et al. (2001)}
\tablefoottext{g}{This work}
\tablefoottext{h}{$ M_{\rm Bar} =M_{\rm Star}+M_{\rm HI}+M_{\rm H_2}+M_{\rm tail} $}
}

\label{Table1}
\end{table}

\section{The galaxy} \label{galaxy}
In spite of its SBab classification, the morphology of NGC 4848 is rather 
irregular, being composed of a central nuclear source (harboring an active galactic nucleus (AGN) [NII]$\lambdaup 6584\AA$/H$\alpha$=0.64, 
see Gavazzi et al. 2011) with red ($g-i$=1.5 mag) color. At a $5-10$ arcsec projected  
radial distance from the nucleus, a bright, blue, and clumpy ring of active HII regions 
dominates the galaxy morphology. Further out, the light profile 
falls-off exponentially, as is typical of a disk. Given the irregular shape of the galaxy, a simple 
(bulge+disk) model  cannot fit the measured light profile in any band.
Therefore, a non-parametric classification method has to be used, such as the CAS system of Conselice (2003).

Based on the  $\rm H\alpha$ and $r$-band images obtained in 2012 and on B and H-band images, 
we derived the concentration, asymmetry and clumpiness (CAS) parameters of NGC 4848 (see Table \ref{CAS}).
We remind the reader that a theoretical $r^{1/4}$ law has $C=5.2$, while an exponential disk has $C=2.7$. 
Moreover, a symmetric and smooth galaxy has $A=0$, $S=0$. Typical values for Sa-Sb galaxies in the $r$ band are 
$C=3.9 \pm 0.5$, $A=0.07 \pm 0.04$, and $S=0.08 \pm 0.08$ (Conselice 2003), which are consistent with the values of
$C=3.0 \pm 0.5$, $A=0.11 \pm 0.05$, and $S=0.10 \pm 0.10$ found by Fossati et al. (in preparation) for a larger sample of 360 galaxies in the Coma
Supercluster in the $r$ band. The same set of parameters were derived in the NET $\rm H\alpha$ frames, yielding 
$C=2.5 \pm 0.9$, $A=0.28 \pm 0.11$, and $S=0.65 \pm 0.26$ for Sa-Sb.
Therefore, NGC 4848 is significantly more asymmetric and 
clumpy and less concentrated than typical Sab galaxies, especially in bands that more sensitively trace 
the presence of young stars, consistently suggesting that some morphological peculiarity has been recently 
($<1$ Gyr) induced by the interaction of the galaxy with its surrounding environment.

\begin{table}[!h]
\begin{center}
\normalsize
\caption{The concentration, asymmetry, and clumpiness (CAS) parameters of the galaxy NGC 4848 in four bands.}
\begin{tabular}{l c c c c}
\hline
\hline
Band & $C$ & $A$  & $S$  \\
\hline
$\rm{H\alpha}$    &   1.76   &   1.05  & 1.75	   \\	      
B                 &   1.83   &   0.45  & 1.06	   \\	      
$r$               &   2.50   &   0.30  & 0.57	   \\	      
H                 &   3.06   &   0.18  & 0.35	   \\	      
\hline                                                                                                      
\hline                                                     
\end{tabular}                                              
\label{CAS}
\end{center}
\end{table}

\subsection{The SFR from the $\rm H\alpha$ luminosity}

\begin{figure*}[t]
\centering
\includegraphics[scale=0.30]{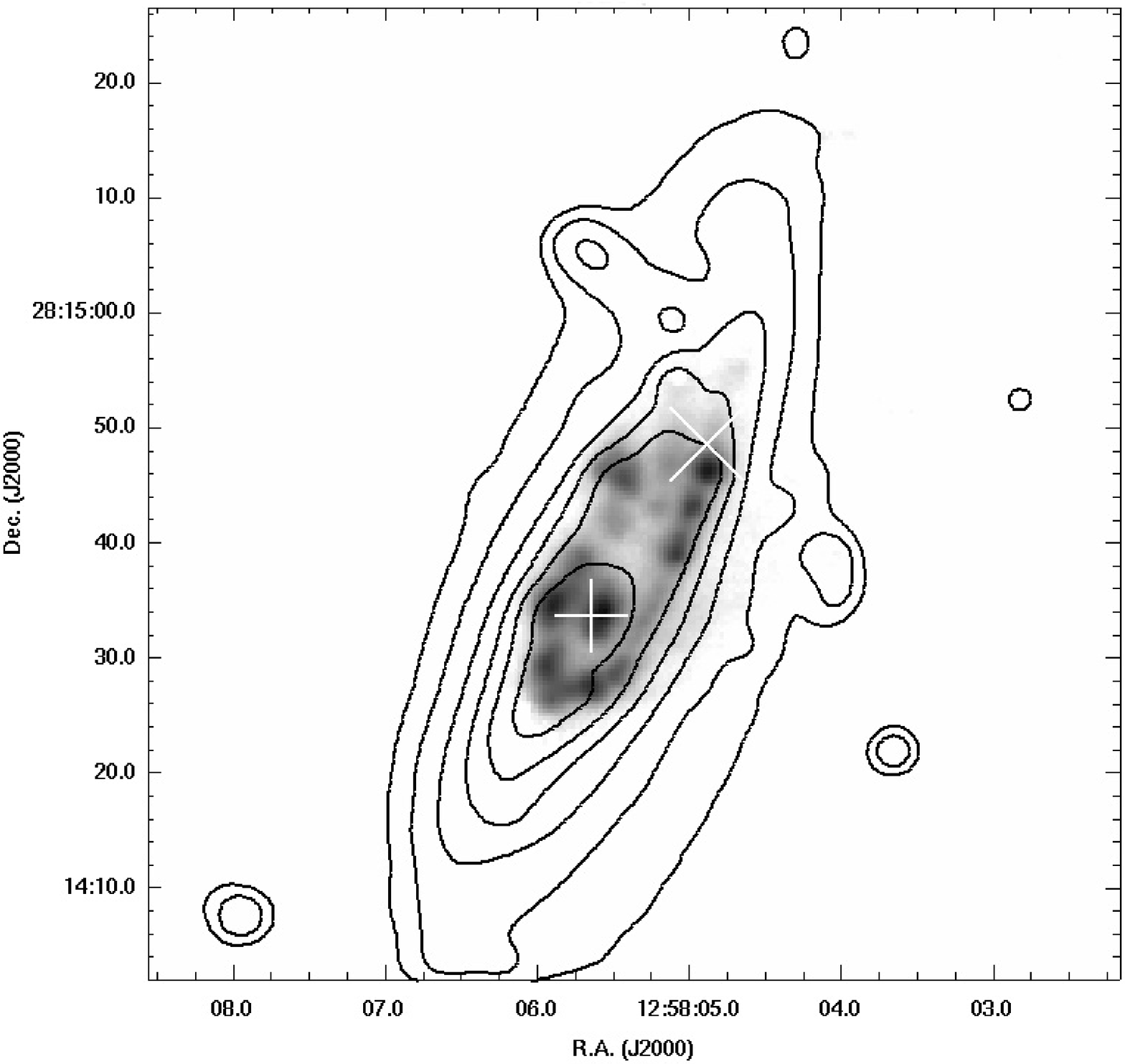}
\includegraphics[scale=0.24]{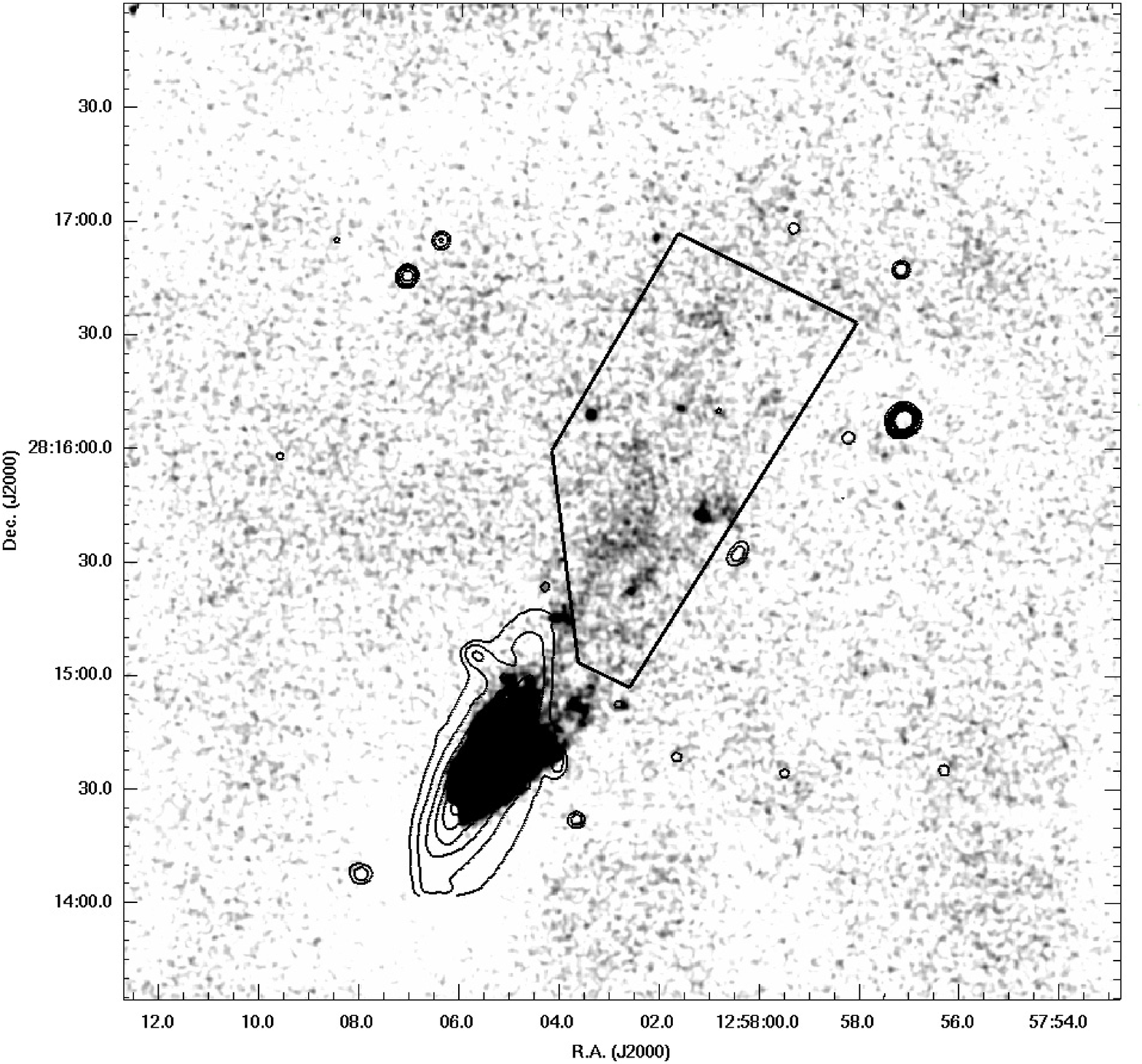}
 \caption{Left panel: Grayscale representation of the inner structure of the NET image of NGC 4848 at full resolution, 
 with contours of the OFF-band image. The white cross marks the center of the molecular hydrogen
given by Vollmer et al. (2001), while the white X marks the center of the HI given by Bravo Alfaro et al. (2001). 
Right panel: High-contrast representation of the Gaussian-smoothed (three-pixel-kernel) NET image to highlight the 
emission in the tail, with contours of the OFF-band image. The black box marks the region where the flux in the tail was computed.}
\label{NET2}
\end{figure*}

To make a quantitative estimate of the $\rm H{\alpha}+[NII]$ flux associated with NGC 4848, we preferred to use the new SPM image alone
to avoid uncertainties caused by to resampling that occurs when the image is stacked with the existing INT observation. 
The $\rm H{\alpha}+[NII]$ flux within the optical extent of NGC 4848 was found to be $\log F\rm(H\alpha+[NII]) = -12.47 \pm 0.03~\rm erg~cm^{-2}~sec^{-1}$.  
Before converting this flux into a star-formation rate (SFR), the $\rm H\alpha$ flux must be corrected for
Galactic extinction, contamination due to [NII]$\lambdaup\lambdaup 6548,6584\AA$, internal extinction, and dust extinction.
(see Boselli et al. 2009, Gavazzi et al. 2012). 
Given the high Galactic latitude of the Coma cluster, extinction due to the Milky Way (Schlegel et al. 1998) is negligible 
(the dust-corrected value is $\log F\rm(H\alpha+[NII])_{MW} =-12.46 \pm 0.03~~\rm erg~cm^{-2}~sec^{-1}$). 
The deblending from [NII] was computed using the drift-scan spectrum of the galaxy taken from Gavazzi et al. 
(2004). The corrected flux became $\log F\rm(H\alpha)_{MW;DB} = -12.68 \pm 0.03~~\rm erg~cm^{-2}~sec^{-1}$.
The flux was further corrected for internal extinction using the Balmer decrement (as in Boselli et al. 2012, submitted),
which was determined again with the drift-scan spectrum to yield $\log F\rm(H\alpha)_{MW;DB;AA} = -12.11 \pm 0.03~~\rm erg~cm^{-2}~sec^{-1}$.
The final correction for dust absorption in HII regions assumes that only a fraction $f$ of the Lyman continuum photons
contribute to the ionization of the atomic hydrogen. Following Boselli et al. (2009) and assuming a Salpeter IMF, $f =0.77$
we obtained $\log F\rm(H\alpha)_{MW;DB;AA;EX} = -11.99 \pm 0.03~~\rm erg~cm^{-2}~sec^{-1}$.
Conversely, the photon escape fraction was considered to be negligible (Boselli et al. 2009).
The corrected  $\rm H\alpha$ luminosity was thus $L_{\rm H\alpha}= 1.1 \pm 0.2 \times 10^{42} ~\rm erg~ sec^{-1}$. 
According to Kennicutt (1998), the unobscured star-formation rate is given by 
$SFR_{\rm H\alpha} [M_{\odot} ~yr^{-1}]  = 7.9 \times 10^{-42} \times L_{\rm H\alpha} [\rm erg~ sec^{-1}]$, thus
$SFR_{\rm H\alpha} \rm = 8.8 \pm 0.2 ~M_{\odot} ~yr^{-1}$.
The bulk of the star formation takes place in bright HII regions distributed
along the central ring-like pattern, as discussed above (see Fig. \ref{NET2} Left).

\subsection{The SFR from the UV luminosity}

Using Kennicutt (1998) $SFR_{\rm UV} [M_{\odot} ~yr^{-1}] = 1.4 \times 10^{-28} \times L_{\rm UV} [\rm erg~ sec^{-1}~Hz^{-1}]$, where
$L_{\rm UV}$ is obtained following Boselli (2011) from the GALEX far-ultraviolet (FUV) magnitude of 16.29 mag (Martin et al. 2005). 
The UV flux is particularly affected by dust extinction and the correction (A(FUV)=2.01 mag) is obtained
following Cortese et al. (2008) assuming that there is a balance between the absorbed UV photons and the total FIR (3-1100 $\rm \mu m$) energy. 
From the corrected UV magnitude, we obtained $SFR_{\rm UV} \rm = 10.7  ~M_{\odot} ~yr^{-1}$.

\subsection{The SFR from the FIR luminosity}

According to Helou et al. (1985), the FIR 8-1000 $\rm \mu m$ luminosity is obtained by combining the
12, 25, 60, and 100 $\rm \mu m$ flux from IRAS (Neugebauer et al. 1984). Since NGC 4848 is undetected in the IRAS 12 $\rm \mu m$
band, we used the WISE (Wright et al. 2010) 12 $\rm \mu m$ flux. We obtained $F_{\rm IR}= 1.51 \times 10^{-13}~ \rm W m^{-2}$.
Using the Kennicutt (1998) calibration $SFR_{\rm IR} [M_{\odot} ~yr^{-1}] = 4.5 \times 10^{-44} \times L_{\rm IR} [\rm erg~ sec^{-1}]$,  
we obtained $SFR_{\rm IR} \rm = 7.4  ~M_{\odot} ~yr^{-1}$.

\subsection{The SFR from the radio continuum luminosity}

The estimate of the SFR from the radio continuum (RC) is based on the 1415 MHz flux (21.8 mJy) from NVSS (Condon et al. 2002) corrected for 
the nuclear contribution (7.1 mJy) from FIRST (Becker et al. 1994). Following the Bell et al. (2003) calibration, 
$SFR_{\rm RC}[\rm{  M_{\odot} ~yr^{-1}}] = 5.52 \times 10^{-22}\times L_{\rm RC} [\rm W~Hz^{-1}]$,
we obtained $SFR_{\rm RC} \rm = 8.8  ~M_{\odot} ~yr^{-1}$.

Assuming that the star formation is constant during the last $10^{7}(\rm{ H\alpha})-10^{8}(\rm UV,FIR,RC)$  yr (Iglesias-P{\'a}ramo et al. 2004),
the average star-formation rate becomes $SFR = 8.9 \pm 1.1 \rm ~M_{\odot} ~yr^{-1}$, corresponding to a specific star-formation 
rate (SSFR) of $1.9 \times 10^{-10} \rm ~yr^{-1}$ .
This is more than a factor of ten higher than the SSFR of galaxies of similar mass and Hubble type. In conclusion, the galaxy appears 
to be undergoing a starburst phase possibly triggered by the interaction with the cluster environment (as claimed by Bekki \& Couch 2003).

\section{$\rm H\alpha$ tail parameters}
\label{trail}

The flux estimate of the low-brightness tail requires a careful assessment of the quality of the  
flat-fielding and background subtraction. For extended sources, 
the dominant source of error is associated with the background variations on scales similar
to those of the source. We measured the residual background in several $\rm 20 \times 20~arcsec^2$ regions 
(comparable with the extension of the trail) within the field and determined that the median of the 
residual background on this scale corresponds to 10 \% of the background rms. From this figure,
we estimated the uncertainty in the the H$\alpha$  flux in the tail (see Gavazzi et al. 2012).

At the distance of the Coma cluster, the total extent of the tail  of $2.25 \times 0.66$ arcmin$^2$ 
corresponds to $62.5  \times 18.5$ kpc. To estimate the total flux in the tail, we integrated the counts 
above the 2-$\sigma$ level in the full-resolution SPM image alone inside a polygonal region  
of  $2.25 \times 0.66$ arcmin$^2$, excluding the galaxy itself and the compact knots 
(see below) (the polygonal region is sketched in Figure \ref{NET2}). 
We obtained $\log F_{\rm tail} = -14.40 ~(-14.66) \pm 0.12 ~~\rm erg~cm^{-2}~sec^{-1}$.
Hereafter, the value in parenthesis is obtained by correcting for [NII]$\lambdaup\lambdaup 6548,6584\AA$
contamination when a [NII]/H$\alpha$ ratio of 0.62 is assumed, as for the main galaxy. 
\footnote{The presence of a central AGN might lead to an overestimate of the [NII]/H$\alpha$ ratio.
However, Yoshida et al. (2012) found that this ratio in the filaments 
associated with IC4040, another giant spiral galaxy in the Coma cluster, 
is that of a LINER-like spectrum.}
No correction for internal extinction was applied to the tail. 

The line intensity of the low-brightness tail was obtained using
\begin{equation}
I_{\rm{H\alpha}}=\frac{F_{\rm{tail}}}{h\nu_{\rm {H\alpha}}\times \Omega}=0.13~(0.07) ~\rm Rayleigh
\end{equation}
(1 Rayleigh = $\rm 10^6/4\pi~photons~cm^{-2}~s^{-1}~sr^{-1}$). 
Following Spitzer (1978), the H$\alpha$ line radiation from an optically thin nebula at $T=10^4$ K
is customarily expressed in terms of the emission measure (EM). Adopting an 
``effective recombination coefficient'' of $\alpha_{32}=1.17 \times 10^{-13}\rm cm^3 s^{-1}$ (Osterbrock 1989), 
we derived $\rm 1 Ry_{H\alpha} = 2.77 ~EM_{H\alpha} $.\\
In the case of NGC 4848, EM amounts to 0.36 (0.19) $\rm cm^{-6}~pc$.

Assuming that the tail has cylindrical symmetry,
this implies a mass of ionized hydrogen of
$5.0\times 10^9$ ($3.6 \times 10^9$) $\rm M_{\odot}$, thus a plasma density of 
$4.3 \times 10^{-3}$ ($3.2 \times 10^{-3}$) $\rm cm^{-3}$. If the filling factor is $<1$, the values obtained
should be taken as an upper limit to the mass and a lower limit to the density.
The plasma in the tail must have originally consisted of neutral hydrogen, therefore
a rough estimate of the total gas loss from the galaxy can be derived using the HI deficiency parameter
defined by Giovanelli \& Haynes (1985). 
The galaxy contains $2.2 \times 10^9$  $\rm M_{\odot}$ (Gavazzi 1989), corresponding to  
an HI deficiency parameter of 0.46, implying that $4.1 \times 10^9$  $\rm M_{\odot}$ of gas has been left behind in the tail.
This estimate is consistent with the mass previously found in the ionized gas.

We assumed a transit velocity through the cluster of $\rm \sim 1330~km~s^{-1}$, as derived 
from $\sqrt 2 \sigma_{vel}$, (where we also assumed that the cluster can be modeled as an isothermal sphere,
as in Cayatte et al. 1994), where $\sigma_{vel}=940 ~\rm km~s^{-1}$ for A1656 (Gavazzi et al. 2010).
From the length of the tail, we then derived that the ionized material survived some $10^{7.6}$ yr.
The recombination time $\tau_r=1/N_e \alpha_{32}$, is $10^{7.8} (10^{7.9}$) yr, i.e. consistent with the survival time.
Although recent simulations favor a scenario where the gas is heated in situ to H$\alpha$ emitting temperatures
and the eddies along the tail contain sufficient turbulent energy to currently sustain
its ionization (Tonnesen \& Bryan 2010), it cannot be excluded that part of the gas in the tail was ionized by the galaxy
HII regions or AGN. 

In addition to the ionized diffuse gas, the tail of NGC 4848 harbors some
compact star-forming regions (possibly HII regions) that we label K1-3 in Figure \ref{NET}. They are
visible on the  $r$ band image and they coincide with compact features in the FUV image by Smith et al. (2010). 
Their uncorrected $\rm H{\alpha}+[NII]$ flux is
$3.6\times 10^{-16}$,  $2.1\times 10^{-16}$, and $7.4\times 10^{-16}~\rm erg~cm^{-2}~sec^{-1}$ respectively.
This corresponds to a luminosity of $4\times 10^{38}$, $2.3\times 10^{38}$, and $8\times 10^{38}$ $\rm erg~ sec^{-1}$,
which are typical of HII regions, and represent $\sim 25 \%$ of the total luminosity in the tail.
Similar ``fireballs'' are found in the wakes of other stripped galaxies, such as GMP3016 and GMP4060 in the Coma cluster
(Yagi et al. 2010), or either VCC 1217 (Fumagalli et al. 2011) and VCC 1249 (Arrigoni Battaia et al. 2012) in the 
Virgo cluster. These authors claim that these HII regions were recently born in situ out of stripped material,
as predicted by some hydrodynamical simulations (e.g. Kapferer et al. 2009, Tonnesen \& Bryan 2012).

Boissier et al. (in preparation) claim the absence of in situ star formation behind galaxies in the Virgo
cluster displaying  HI (Chung et al. 2007) but no H$\alpha$ tails. 
As explicitly mentioned by Tonnesen \& Bryan (2010, 2012), the pressure by the IGM (e.g. density, velocity dispersion) is crucial for determining the formation of
relevant eddies that might increase the IGM density enough to provide the cooling conditions driving the formation of molecular clouds,
thus of stars.  Such conditions are met in the Coma cluster and near the center of Virgo, but not further out where most of the Boissier et al. objects are located.

\section{Discussion}

There is unanimous consensus (Yagi et al. 2010, Yoshida et al. 2012) that the morphological disturbances suffered by 
several late-type galaxies in the Coma cluster 
are caused by the dynamical interaction with the IGM, namely by the ram pressure mechanism  (Gunn \& Gott 1972),
including NGC 4848 (Gavazzi 1989, Vollmer et al. 2001).
The question is how long this process has been acting for and whether the galaxy is infalling into the cluster
for the first time or, as maintained by Vollmer et al. (2001), it entered the cluster environment 
about 1 Gyr ago and already crossed the cluster center 400 Myr ago. 

From the length of the tail, we inferred that the galaxy motion takes place primarily in the plane of the sky, 
probably at $V \rm \sim 1330~km~s^{-1}$  ($\sqrt 2 \sigma_{vel}$). Unfortunately, we cannot directly
measure this velocity since from the redshift we can infer only the component along the line of
sight ($V_{//}$). Subtracting from the heliocentric velocity of NGC 4848 (Zabludoff et al. 1993), the
mean velocity of Coma cluster galaxies (Gavazzi et al. 2010), we obtained $V_{//}=257~\rm km~s^{-1}$. 

There are two possibilities: that the orbit is circular,  with its angular momentum
in the plane of the sky, or that it is radial. In the first case, the ratio of the velocity along 
the line of sight to the tangential velocity $V_{//}/V = 257/1330 = 0.19$
gives the angle between $V_\perp$ and $V$. Assuming that the present projected radial distance of NGC 4848
from the cluster center is 0.75 Mpc, we conclude that the radius of the circular orbit is 4 Mpc,
i.e. exceeds the virial radius of Coma by almost a factor of two. 
We are much more likely to conclude, however, that the orbit is radial or at least highly eccentric, 
and that the present motion of NGC 4848
is toward the center of the cluster at $V \rm \sim 1330~km~s^{-1}$, as suggested by the direction of the 
H$\alpha$ tail pointing in the opposite direction. 

More evidence of infall into the Coma cluster is provided by the velocity distribution of late-type galaxies (LTG)
which in general hardly follow a Gaussian distribution but one that is skewed toward either lower or higher velocities
than the overall mean velocity 
(Biviano et al. 1997, Boselli \& Gavazzi 2006). The velocity dispersion profile of the LTGs in clusters is found 
to be consistent with orbits more radial than those of early-type galaxies (ETG),
providing a picture in which possibly all spirals have not yet crossed the virialized cluster core, 
and may even be on a first (infall) approach toward the central, high-density region.
 
In the ram pressure scenario, the amount of stripped gas can be computed from the equilibrium between
the gravitational binding energy and the dynamical pressure. The calculation requires the density profile of the
gas in the IGM. Following Cavaliere \& Fusco-Femiano (1978), the density radial profile is well-approximated by an
isothermal sphere
\begin{equation}
\rho_{\rm{IGM}}=\rho_0\left[1+\left(\frac{r}{r_c}\right)^2\right]^{-3\beta/2},
\end{equation}
where $\rho_0$ represents the central density and $r_c$ the scale-length.
Using the parameters estimates of Mohr et al. (1999) for the Coma cluster (all quantities having been recomputed for $\rm H_0=73~km~s^{-1}~Mpc^{-1}$)
namely $\rho_0=6.3\times 10^{-27} ~\rm g~ cm^{-3}$, $r_c=0.26 \rm ~Mpc$, and $\beta=0.7$, we computed that
at the present position of NGC 4848 (0.75 Mpc from the center) the IGM density is
$\rho_{0.75}=6\times 10^{-28} ~\rm g~ cm^{-3}$.

The radius at which ram pressure becomes efficient can be estimated as (Domainko et al. 2006)
\begin{equation}
R_{\rm{strip}} = 0.5 R_{\rm{HI}} ln \left (\frac{GM_{\rm{star}}M_{\rm{HI}}}{v^2 \rho_{\rm{IGM}} 2 \pi R_{\rm{star}}^2
R_{\rm{HI}}^2} \right),
\end{equation}
while the stripped mass is 
\begin{equation}
M_{\rm{strip}} = M_{\rm{HI}} \left(\frac{R_{\rm{strip}}}{R_{\rm{HI}}}+1\right) exp\left(-\frac{R_{\rm{strip}}}{R_{\rm{HI}}}\right).
\end{equation}
In this calculation, we assumed exponential profiles for both the stars and the interstellar gas, 
with $R_{\rm star}=5.0$ kpc (computed on the H band image) and $R_{\rm HI}=1.8 \times R_{\rm star}=9.0$
kpc for the HI disk (Boselli \& Gavazzi 2006).
This yielded $R_{\rm{strip}} = 8.5$ kpc and $M_{\rm{strip}} = 4.7 \times 10^9 \rm ~M_\odot ~\sim 0.75 ~M_{HI~orig}$.
This is in remarkably good agreement with the mass of stripped gas ($3.6\times 10^9 \rm ~M_\odot$) computed in \S \ref{trail}, which in turns
is consistent with the missing mass of atomic hydrogen ($4.1\times 10^9 \rm ~M_\odot$) computed from 
the HI deficiency parameter.

Using the ram pressure simulation by Kapferer et al. (2009) (the case with $V_{rel}=1000 \rm ~km~s^{-1}$) and assuming 
the IGM density computed above ($\rho_{0.75}=6\times 10^{-28} ~\rm g~ cm^{-3}$), we derived that 65 \% of the original HI content
is stripped in about 200 Myr. In spite of the different conditions found in the Virgo cluster with respect to the Coma Cluster,
a similarly short timescale ($\sim 100 \rm ~Myr$) is reported by Boselli et al. (2006) for an almost complete ablation of the atomic gas from NGC 4569. 
This time is significantly shorter than the crossing time in the Coma cluster ($1.6\times 10^9$ yr,
Boselli \& Gavazzi 2006), supporting the conclusion that NGC 4848 is on its first passage through the cluster core. 
This time would be sufficient to remove most of its gas, in contradiction to the 1 Gyr timescale proposed by Vollmer et al. (2001).

\acknowledgements

We acknowledge useful discussions with Massimo Dotti.
We thank an anonymous referee for providing us 
with constructive comments and suggestions.
This work made extensive use of GoldMine, 
the Galaxy On Line Database (http://goldmine.mib.infn.it).
G. Gavazzi acknowledges financial support from Italian MIUR PRIN contract 200854ECE5.
The present study could not have been conceived without the DR7 of SDSS. 
Funding for the Sloan Digital Sky Survey (SDSS) and SDSS-II has been provided by the 
Alfred P. Sloan Foundation, the Participating Institutions, the National Science Foundation, 
the U.S. Department of Energy, the National Aeronautics and Space Administration, 
the Japanese Monbukagakusho, and 
the Max Planck Society, and the Higher Education Funding Council for England. 
The SDSS Web site is \emph{http://www.sdss.org/}.
The SDSS is managed by the Astrophysical Research Consortium (ARC) for the Participating Institutions. 
The Participating Institutions are the American Museum of Natural History, Astrophysical Institute Potsdam, 
University of Basel, University of Cambridge, Case Western Reserve University, The University of Chicago, 
Drexel University, Fermilab, the Institute for Advanced Study, the Japan Participation Group, 
The Johns Hopkins University, the Joint Institute for Nuclear Astrophysics, the Kavli Institute for 
Particle Astrophysics and Cosmology, the Korean Scientist Group, the Chinese Academy of Sciences (LAMOST), 
Los Alamos National Laboratory, the Max-Planck-Institute for Astronomy (MPIA), the Max-Planck-Institute 
for Astrophysics (MPA), New Mexico State University, Ohio State University, University of Pittsburgh, 
University of Portsmouth, Princeton University, the United States Naval Observatory, and the University 
of Washington.\\

\smallskip

\end{document}